\documentclass[prl,nofootinbib,showpacs,preprint]{revtex4}
\usepackage{subfigure}
\usepackage{amsmath}
\usepackage{amsfonts}
\usepackage{amssymb}
\usepackage[dvips]{graphicx}

\usepackage[colorlinks]{hyperref}
\usepackage[usenames,dvipsnames]{color}
\usepackage[applemac]{inputenc}
\usepackage[T1]{fontenc}
\newcommand{\half}{\tfrac{1}{2}}
\usepackage{bbm}
\def\beq{\begin{equation}}
\def\eeq{\end{equation}}
\def\bea{\begin{eqnarray}}
\def\eea{\end{eqnarray}}
\def\ba{\begin{array}}
\def\ea{\end{array}}

\def\part{\partial}
\def\tfrac#1#2{{\textstyle{#1\over #2}}}
\def\half{\tfrac{1}{2}}
\def\haf#1{\tfrac{#1}{2}}

\begin{document}

\preprint{UdeM-GPP-TH-16-247}
\preprint{arXiv:1601.00475[hep-th]}
\title{Solitons in an effective theory of CP violation}
\author{N. Chandra$^{1,2,3}$}
\email{nitinc@imsc.res.in}
\author{M. B. Paranjape$^2$}
\email{paranj@lps.umontreal.ca}
\author{R. Srivastava$^{1,2,3}$}
\email{rahuls@imsc.res.in}
%\author{Sachindeo Vaidya$^{1}$}
%\email{vaidya@cts.iisc.ernet.in}

\affiliation{$^1$  Center for High Energy Physics, Indian Institute of Science, Bangalore 560012, India}
\affiliation{$^2$Groupe de physique des particules, D\'epartement de
physique, Universit\'e de Montr\'eal, C.P. 6128, succ. centre-ville,
Montr\'eal, Qu\'ebec, CANADA, H3C 3J7 }
\affiliation{$^3$ Institute of Mathematical Sciences, IV Cross Road, CIT Campus,
Taramani, Chennai 600113
Tamil Nadu, India.}

\begin{abstract}
We study an effective field theory describing  CP-violation in a scalar  meson sector.  We write the simplest interaction that we can imagine,  $${\cal L}\sim \epsilon_{i_1\cdots i_5}\epsilon^{\mu_1\cdots\mu_4}\phi_{i_1}\partial_{\mu_1}\phi_{i_2}\partial_{\mu_2}\phi_{i_3}\partial_{\mu_3}\phi_{i_4}\partial_{\mu_4}\phi_{i_5}$$ which involves 5 scalar fields. The theory describes CP-violation only when it contains scalar fields representing mesons such as the $K^*_0$, sigma, $f_0$ or $a_0$.  If the fields represent pseudo-scalar mesons, such as  B, K  and $\pi$ mesons then the Lagrangian describes anomalous processes such as $KK\to \pi\pi\pi$.     We speculate that the field theory contains long lived excitations corresponding to $Q$-ball type domain walls expanding through space-time.  In an 1+1 dimensional, analogous, field theory we find an exact, analytic solution corresponding to such solitons.  The solitons have a U(1) charge $Q$, which can be arbitrarily high, but oddly, the energy behaves as $Q^{2/3}$ for large charge, thus the configurations are stable under disintegration into elementary charged particles of mass $m$ with $Q=1$.   We also find analytic complex instanton solutions which have finite, positive Euclidean action.  
\end{abstract}

\maketitle
\section{Introduction}
CP violation is predicted by the standard model \cite{burgess2006standard}, and exists because of the Kobayashi-Maskawa mass matrix \cite{kobayashi1973cp} which crucially involves and mixes three flavours of quarks.  However, the CP-violation in the standard model is woefully inadequate to describe the baryon asymmetry of the universe \cite{sakharov1967pis,Kuzmin:1970nx,Weinberg:1979bt}.   In this letter we  look for new non-perturbative sources of CP violation within the context of the standard model.  Solitons and instantons, classical field configurations in general, are understood to contribute to quantum amplitudes in a non-perturbative dependence on the coupling constant \cite{Coleman:1978ae}.  Here we look for solitons-like configurations in an effective theory of mesons.  Such an effective theory  would arise within a low energy description of the dynamics of the mesons in the standard model.  

CP-violation could be modelled, in a possible effective description, by the Lagrangian containing five real scalar (not pseudo-scalar) fields $\phi_i, i=1\cdots 5$ representing the various mesons, with a CP violating interaction term:
\beq
{\cal L}=\half \left(\partial_\mu\phi_i\partial^\mu\phi_i -m_i^2\phi_i^2\right) +\haf\lambda\epsilon_{i_1\cdots i_5}\epsilon^{\mu_1\cdots\mu_4}\phi_{i_1}\partial_{\mu_1}\phi_{i_2}\partial_{\mu_2}\phi_{i_3}\partial_{\mu_3}\phi_{i_4}\partial_{\mu_4}\phi_{i_5}\label{1}
\eeq
where a sum over repeated latin indices from $1,\cdots ,5$ and a sum over repeated greek indices from $0,\cdots ,3$ is understood.  Such interactions have been considered before in the context of Bi and Multi Gallileon theories \cite{Padilla:2010ir,Padilla:2010de}.  This Lagrangian is CP violating  if the fields are taken even under time reversal.   Lorentz invariance implies the CPT theorem \cite{streater2000pct}, hence CP violation is the same as time reversal violation. The interaction in the Lagrangian \eqref{1} is odd under time reversal.  It is easy to imagine that there are other terms in the Lagrangian that give rise to CP conserving interactions between the mesons and require the fields be time reversal even.

The pseudo-scalar B mesons decay to lighter hadronic mesons through flavour changing, charged current, weak leptonic decays that contain CP violating channels \cite{Durieux:2015zwa, Aaij:2014qwa, gronau1990isospin}.   The decay$B\rightarrow 2\,K\,2\,\pi$,  is of great specific interest in the experiment LHCb that is going on at the present time at the accelerator at CERN \cite{alves2008lhcb} .  The interaction in \eqref{1} cannot describe such decays as it is not CP-violating for pseudo-scalar meson fields.  However, the interaction in \eqref{1} appears as the lowest order term in the expansion of the Wess-Zumino-Novikov-Witten \cite{wess1971consequences,Witten:1983tw,Novikov:1982ei} that must be added to the usual Skyrme \cite{skyrme1961non,skyrme1962unified,gisiger1998recent} model.  The interaction in \eqref{1} then describes anomalous processes such as $KK\to \pi\pi\pi$ which are of course allowed in QCD but absent in the usual Skyrme model without the WZNW term \cite{Witten:1983tw,Witten:1983tx, witten1979baryons}.  

Consider the ansatz
\bea
\phi_1+i\phi_2&=&f(r)e^{i\omega t}\\
(\phi_3,\phi_4,\phi_5)&=&g(r)\hat r(\theta,\varphi)
\eea
with a mass $m$ for the fields $\phi_1$ and $\phi_2$, and zero mass for the remaining fields.
This ansatz yields the equations of motion:
\bea
-\omega^2 f-(1/r^2)(r^2f')'+m^2f+60\lambda g'g^2f/r^2&=&0\\
-(1/r^2)(r^2g')'+60\lambda\omega g'gf^2/r^2&=&0
\eea
We imagine the existence of localized, finite energy solutions to these equations of motion.  The fields $f(r)$ and $g(r)$ both vanish at the origin and stay negligible until they reach a certain radius $R$.  Here they exhibit non-trivial behaviour, we presume $f$ has a small, positive bump while $g(r)$ interpolates to +1, and for larger $r$, $f(r)\to 0$ while $g(r)\sim 1$, although, it could well be that both fields vanish at spatial infinity.  Such a configuration could be of finite energy, and depending on what other terms might be added to the Lagrangian.  Usually the non-trivial dependence of the fields $\phi_3,\phi_4\phi_5$ at $\infty$ would correspond to infinite energy, however we speculate that this is not the case.  In any case, in the cosmological context, infinite energy solitons are not prohibited \cite{vilenkin2000cosmic}, for example, global strings are permitted. The configurations could be stable or unstable to  expansion or contraction, however, we  expect the configurations to be generally long lived.  In that way, they could give rise to non-perturbative contributions to CP-violating processes.   The analysis of this 3+1 model will be left to a future publication. 

Our intuition is gleaned from the study of an analogous 1+1 dimensional  model, where surprisingly, we find exact, analytic soliton and instanton solutions.   Our analysis  gives plausibility to the possibility that the 3+1 dimensional model contains soliton solutions and even instantons.   The 1+1 dimensional instantons have a nontrivial winding at infinity, but the action is finite, which lends credence to our impression that the analogous 3+1 dimensional solutions of finite energy, would also exist.    Their higher dimensional analogs would be infinite domain wall type solitons, or  closed (spherical) domain walls giving rise to spherical solitons.  The existence and stability of the 3+1 dimensional configurations is not studied in this paper. 
\section{Minkowski 1+1 dimensional model}
The analog of the model \eqref{1} in 1+1 dimensions contains three real scalar fields.  We will write the Lagrangian for arbitrary masses, but we will specialize when we solve the equations of motion.
\subsection{Action and the Equations of Motion}
We will study the equations of motion corresponding to the Lagrangian density given by
\begin{equation} \label{L}
\mathcal{L} = \frac{1}{2} \left[(\partial_\mu \phi_i)(\partial^\mu \phi_i) - m_i^2 \phi_i^2 + \lambda \epsilon_{ijk}\epsilon^{\mu\nu}\phi_i(\partial_\mu \phi_j) (\partial_\nu \phi_k)\right]
\end{equation}
where summations over repeated indices are to be understood. $\mu,\nu = 0,1$ and $i,j,k = 1,2,3$.
The equations of motion are simply
\begin{equation}
\partial_\mu\partial^\mu \phi_i + m_i^2 \phi_i - \frac{3}{2}\lambda \epsilon_{ijk} \epsilon^{\mu\nu} (\partial_\mu \phi_j) (\partial_\nu \phi_k) = 0
\end{equation}
for  $i=1,2,3$. 

\subsection{Energy}
The Lagrangian \eqref{L} is invariant under the time translation giving rise to energy conservation.  The interaction term being linear in time derivatives,  does not contribute to the Hamiltonian and consequently nor to the energy.  
%The easiest way to see this is via the energy momentum tensor, which would correspond to the variation of the action with respect to the metric, evaluated of course afterwards at the Minkowski metric.  The interaction term is in fact independent of the metric since the anti-symmetric tensor $\epsilon^{\mu\nu}$ becomes the tensor density $\epsilon^{\mu\nu}/\sqrt{g}$ but the $\sqrt g$ exactly cancels against that which is added when we write the volume element in a general covariant fashion, $d^2x\rightarrow d^2x\sqrt g$.  
%The energy momentum tensor is given by
%\beq
%T^\mu_\nu = (\partial^\mu \phi_i)(\partial_\nu \phi_i) - \frac{1}{2} \delta^\mu_\nu (\partial_\rho \phi_i) (\partial^\rho \phi_i) +\frac{1}{2} \delta^\mu_\nu m_i^2 \phi_i^2 
%\eeq
The energy density is given by
\begin{equation} \label{energy_density_phi}
\varepsilon (x) = T^0_0 = \frac{1}{2} \left[ \dot{\phi_i}\dot{\phi_i} + \phi'_i\phi'_i + m_i^2 \phi_i^2\right]
\end{equation}
the total energy obtained upon integration over space.

\subsection{The Case $m_1=m=m_2, m_3=0$}
%The Lagrangian $\mathcal{L}$
%\begin{equation} \label{L_m3=0}
%\mathcal{L} = \frac{1}{2} \left[(\partial_\mu \phi_i)(\partial^\mu \phi_i) - m^2 (\phi_1^2 + \phi_2^2) + \lambda \epsilon_{ijk}\epsilon^{\mu\nu}\phi_i(\partial_\mu \phi_j) (\partial_\nu \phi_k)\right]
%\end{equation}
The equations of motion are:
\bea
\partial_\mu\partial^\mu \phi_1 + m^2 \phi_1 - \haf{3}\lambda \epsilon_{1jk} \epsilon^{\mu\nu} (\partial_\mu \phi_j) (\partial_\nu \phi_k) &=& 0\\
\partial_\mu\partial^\mu \phi_2 + m^2 \phi_2 - \haf{3}\lambda \epsilon_{2jk} \epsilon^{\mu\nu} (\partial_\mu \phi_j) (\partial_\nu \phi_k) &=& 0\\
\partial_\mu\partial^\mu \phi_3 - \haf{3}\lambda \epsilon_{3jk} \epsilon^{\mu\nu} (\partial_\mu \phi_j) (\partial_\nu \phi_k) &=& 0
\eea
%with energy density
%\begin{equation}
%\varepsilon (x)  = \frac{1}{2} \left[ \dot{\phi_i}\dot{\phi_i} + \phi'_i\phi'_i + m^2 (\phi_1^2 + \phi_2^2)\right]
%\end{equation}
The kinetic term and the interaction are invariant under $SO(3)$ iso-rotations, but these are explicitly, softly broken by the mass terms.  In the present case,  $SO(2)$ symmetry is preserved and the Lagrangian is invariant under an iso-rotation between $\phi_1$ and $\phi_2$, $\phi_1 +i\phi_2\rightarrow e^{i\alpha} (\phi_1+i\phi_2) $.The corresponding conserved current is given by
\beq
j^\mu = \phi_1(\partial^\mu \phi_2) - \phi_2 (\partial^\mu \phi_1) + \lambda \epsilon^{\mu\nu} \left[(\phi_1^2+\phi_2^2)(\partial_\nu \phi_3) - \phi_3(1/2)\partial_\nu (\phi_1^2+\phi_2^2)\right].
\eeq
%Explicitly the charge density is
%\begin{equation} \label{charge_density}
%j^0(x) = \phi_1 \dot{\phi}_2-\phi_2\dot{\phi}_1+\lambda\left[\frac{1}{2}\phi_3(\phi_1^2+\phi_2^2)'-\phi'_3(\phi_1^2+\phi_2^2)\right]
%\end{equation}
%and corresponding conserved charge
%\begin{equation} \label{noether_charge}
%Q = \displaystyle{\int_{-\infty}^{+\infty}} dx \, j^0(x).
%\end{equation}
\subsection {Ansatz and exact soliton}
We take the ansatz
\begin{equation} \label{ansatz_phi1}
\phi_1+i\phi_2 = f(x) e^{i\omega (t-t_0)}\quad\quad\phi_3 = g(x)
\end{equation}
which gives the simple, equations of motion 
\begin{equation} \label{fg_1}
(m^2-\omega^2) f - f'' - 3\lambda\omega fg' = 0
\end{equation}
and
\begin{equation}
 - g'' + 3\lambda\omega ff' = 0\label{2}
\end{equation}
Eqn. \eqref{2} integrates trivially as 
\begin{equation} \label{g'}
g' = \frac{3}{2}\lambda\omega f^2 - A
\end{equation}
where $A$ is a constant. 
The energy density in terms of $f$ and $g$ becomes
\begin{equation}
\varepsilon (x) = \frac{1}{2}\left[f'^2 + g'^2 + (m^2 + \omega^2)f^2\right]\label{ed}
\end{equation}
As each term is a positive definite, the finite energy condition for a solitonic solution requires
$f,f',g' \rightarrow 0$ as $x\rightarrow \pm \infty$.
This condition gives $A=0$ and we get
\beq
g' = \frac{3}{2}\lambda\omega f^2.\label{g'}
\eeq
Putting this back in (\ref{fg_1}) we get remarkably, 
\begin{equation} \label{eom_f}
f'' + \frac{9\lambda^2\omega^2}{2}f^3 - (m^2-\omega^2)f = 0
\end{equation}
which is just the non-linear Schrödinger equation \cite{ablowitz1991solitons}, which is trivially integrable.  We can  rewrite the equation as
\begin{equation}
f'' = - \frac{dU(f)}{d f} 
\end{equation}
with
\begin{equation}
U(f) = \frac{9\lambda^2\omega^2}{8}f^4-\frac{m^2-\omega^2}{2}f^2
\end{equation}
%This is the equation of the ``coordinate" $f(x)$ of a ``particle" with unit ``mass", $x$ being the ``time", moving under a ``potential" $U(f)$. 
As the coefficient of $f^4$ in $U(f)$ is always positive there are typically two types of behaviour of $U(f)$ with respect to $f$  for $m^2\le\omega^2$ and  for $m^2\ge\omega^2$.
%\begin{figure*}[h]
%\begin{center}
%\scalebox{0.7}{\includegraphics{Uvsf1.eps}}
%\scalebox{0.7}{\includegraphics{Uvsf2.eps}}
%\end{center}
%\caption{\label{Uvsf<0} The behaviour of ``potential'' $U(f)$ with respect to $f$ for the case $(m^2-\omega^2)\leq0$}
%\caption{\label{Uvsf>0} The behaviour of ``potential'' $U(f)$ with respect to $f$ for the case $(m^2-\omega^2)> 0$}
%\end{figure*}
A finite energy solitonic solution must satisfy $f\rightarrow 0$ as $x\rightarrow \pm \infty$.

\subsubsection{The case $(m^2-\omega^2)\leq0$}
The only  solution for this case is
\begin{equation}
f(x)=0 \hspace{2 cm} \mbox{{\rm for all }} x
\end{equation}
which gives (see (\ref{g'}))
\begin{equation}
g(x) = \mbox{{\rm constant}} = g_0 
\end{equation}
a constant.  Then the three fields become
\begin{equation}
\phi_1 = \phi_2 = 0, \quad\quad  \phi_3 = g_0.
\end{equation}
The energy  for the above configuration is zero. Thus the above configuration represents a vacuum which is degenerate. Different vacua of the theory correspond to different values for the constant $g_0$. This vacuum solution does not contain any $\lambda$-contribution as the $\lambda$-term in the equations of motion vanishes identically for the above!

\subsubsection{The case $(m^2-\omega^2)>0$}
%Using the ``particle'' analogy for the function $f(r)$, it starts from the origin $O$ in the Figure \ref{Uvsf>0} in the far ``past'', slides down in any one direction, crosses the point where the ``potential'' $U(f)$ is minimum, i.e., $B$ or $C$, reaches to the point ($A$ or $D$ respectively) where the ``potential'' $U(f)$ is again zero at some finite time, and then slides back and retraces the path till it reaches the origin $O$ again at far ``future''.  

We can actually solve  Eqn. (\ref{eom_f}) exactly.  Multiplying it by $f'$ and integrating gives
\begin{equation}
\frac{1}{2}f'^2 + \frac{9\lambda^2\omega^2}{8}f^4 - \frac{1}{2}(m^2-\omega^2)f^2 = \mbox{{\rm constant }} = D.
\end{equation}
Again, finite energy requires that the function $f$ and $f'$  vanish at $x\to\pm\infty$, which requires $D=0$ and we get
\begin{equation}
\frac{1}{2}f'^2 + \frac{9\lambda^2\omega^2}{8}f^4 - \frac{1}{2}(m^2-\omega^2)f^2 = 0.
\end{equation}
This can be written as
\begin{equation}
dx =  \frac{df}{\frac{3\lambda\omega}{2}f\sqrt{\frac{4(m^2-\omega^2)}{9\lambda^2\omega^2}-f^2}}
\end{equation}
where we allow $\omega$ to be positive or negative to allow for either sign in the square root that we have taken.
Noting that $\frac{4(m^2-\omega^2)}{9\lambda^2\omega^2}>0$ and integrating we get
\begin{equation}
x =\pm \frac{1}{\sqrt{m^2-\omega^2}} \mbox{ {\rm  sech}}^{-1} \left[\frac{3\lambda\omega f}{2\sqrt{m^2-\omega^2}}\right] + x_0.
\end{equation}
Inverting
\begin{equation}
f = \frac{2\sqrt{m^2-\omega^2}}{3\lambda\omega} \mbox{ {\rm sech}} \left[\sqrt{m^2-\omega^2}\left(x-x_0\right)\right]\label{f}
\end{equation}
Putting this back in (\ref{g'}) and integrating gives
\begin{equation}
g = \frac{2\sqrt{m^2-\omega^2}}{3\lambda\omega} \mbox{ {\rm tanh}} \left[\sqrt{m^2-\omega^2}\left(x-x_0\right)\right] + g_0
\end{equation}
Here $g(x=x_0) = g_0$.  We notice that for $g_0=0$ we find
\beq
\sum_i\phi^2_i=\frac{4(m^2-\omega^2)}{9\lambda^2\omega^2}\label{sumphi2}
\eeq
which is a constant.
\subsection{Energy and charge}
\begin{figure}[h]
\begin{center}
\scalebox{0.7}{\includegraphics{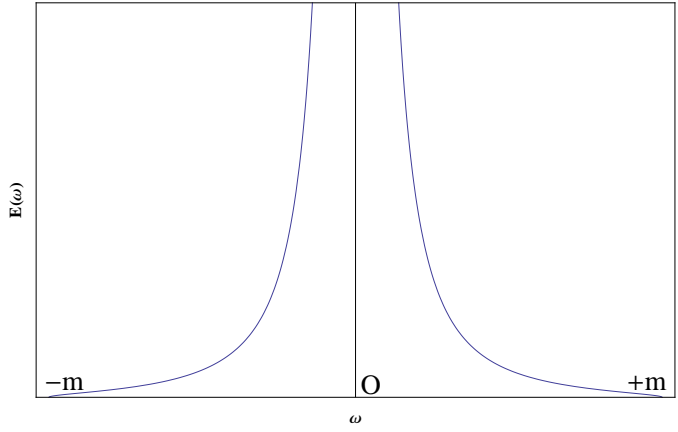}}
\end{center}
\caption{\label{EvsOmega} The behaviour of the Energy of the soliton/anti-soliton with respect to $\omega$}
\end{figure}

The energy for such solutions  is easily calculated from Eqn. \eqref{ed}, we find
\begin{equation}
E = \int_{-\infty}^\infty dx\varepsilon(x)=\frac{8m^2\sqrt{m^2-\omega^2}}{9\lambda^2\omega^2}\label{energy}
\end{equation} 
it's dependence on $\omega$ is shown in Figure \eqref{EvsOmega}. The energy is zero for $\omega=\pm m$ (the degenerate vacua) and increases to infinity as $\omega \rightarrow 0$.

The charge  for the above solution becomes, using the notation $f=\alpha\,{\rm sech}\beta (x-x_0)$, and dropping $x_0$ due to translation invariance, 
\begin{eqnarray}
Q &=&\int dx \left(\omega f^2-\lambda (gff'-f^2g')\right)\nonumber \\
&=&\int dx \left(\omega\alpha^2\,{\rm sech}^2\beta x-\lambda\alpha^3\beta(-\tanh^2\beta x{\rm sech}^2\beta x-{\rm sech}^4\beta x)\right)\nonumber\\
&=&(\omega\alpha^2+\lambda\alpha^3\beta)\int dx\, {\rm sech}^2\beta x=\frac{(\omega\alpha^2+\lambda\alpha^3\beta)}{\beta} 2.
\end{eqnarray}
Replacing for $\alpha$ and $\beta$ from Eqn. \eqref{f} the conserved charge becomes
\begin{equation}
Q = \frac{8\sqrt{m^2-\omega^2}}{9\lambda^2\omega}\left(1+\frac{2(m^2-\omega^2)}{3\omega^2}\right).
\end{equation}
For solitons $\omega$ and hence $Q$ is positive while for anti-solitons they are negative.  We can solve for $\omega$ in terms of $m$ and $E$ from Eqn. \eqref{energy},
\beq
\omega^2=\frac{-64m^4+\sqrt{(64m^4)^2+4(64m^6)81\lambda^4E^2}}{2\cdot81\lambda^4E^2}.
\eeq
For large $E$ this simplifies as
\beq
\omega^2\approx\frac{8m^3}{9\lambda^2E}\left(1+o\left(\frac{m}{E}\right)\right)
\eeq
and then gives
\beq
Q=\left(\frac{1}{3}+\frac{2m^2}{3\omega^2}\right)\frac{8\sqrt{m^2-\omega^2}}{9\lambda^2\omega}\approx \frac{2m^2}{3\omega^2}\frac{8m}{9\lambda^2\omega}\left( 1+o\left(\frac{\omega}{m}\right)\right)\approx 2\lambda\left(\frac{E}{2m}\right)^{3/2}
\eeq
which shows that $E\sim  m \left(\sqrt 2Q/\lambda\right)^{2/3}$ in this limit of large $E$ and hence large $Q$.   This is actually odd for a 1+1 dimensional $Q$-ball.  A general anaylsis \cite{mackenzie2001q} shows that the normal behaviour would be $E\sim Q^{1/2}$.  The actual behaviour that we have found is normally seen in 2+1 dimensional $Q$-balls. Therefore the solitonic configuration is stable compared to $E\sim mQ$ which would be the case for $Q$ perturbative excitations of mass $m$.  

For small $E$ we expand the combination
\beq
\frac{\sqrt{m^2-\omega^2}}{\omega}=\frac{\sqrt{m^2-\omega^2}}{\omega^2}\omega\approx\frac{9\lambda^2 E}{8m^2}m\left(1-\frac{81\lambda^4E^2}{128m^2}+\cdots\right)
\eeq
and then we can express the charge in terms of $m$ and $E$, we get
\beq
Q=\frac{E}{m}\left(1+\frac{1}{6}\cdot\frac{81\lambda^4}{64m^2}E^2\right).
\eeq
This also gives $E<mQ$ which again seems to indicate stability, which is rather surprising, as this would indicate that the perturbative excitations are unstable to forming $Q$ balls, even for individual particles of charge $Q=1$ and mass $m$.  However, at the moment, we  only consider this as in indication, which needs to be verified by numerical calculations.

%which gives, for large $Q$ and $E$, the relation $E\sim m(m/\omega)Q^{1/3}$.  This is to be compared with $E\sim m Q$ which would be valid for $Q$ well separated perturbative excitations, each of mass $m$.  Hence we see that for large $Q$, the soliton is completely stable against disintegrating into $Q$ single particles, each of mass $m$.  This gives the expectation that the solitons, in such cases called a $Q$-ball, is the solution of minimum energy in the sector of charge $Q$.  
\section{Instantons}

\subsection{Action and Equation of Motion}

The Euclidean action is obtained via the analytic continuation $t\to -i\tau$ resulting in $iS_M\to-S_E$, giving
\beq
 S_E  = \frac{1}{2} \, \int d^2x \, \left[(\partial_\mu \phi_i)(\partial_\mu \phi_i) \, + \, m_i^2 \phi_i^2  
\, - \, i \lambda \epsilon_{\mu\nu} \,\epsilon_{ijk} \phi_i(\partial_\mu \phi_j) (\partial_\nu \phi_k) \right]
\label{euclideanaction}
\eeq
where indicies are simply written below as the Euclidean metric is the identity matrix, $g_{\mu\nu}=\delta_{\mu\nu}$.  It is important to note that the interaction term remains imaginary in Euclidean space, this is an example of a complex action, and the corresponding non-trivial solutions to the equations of motion may not be real \cite{alexanian2008path}.   The equation of motion for field $\phi_i$ becomes
\beq
 \partial_\mu \partial_\mu \phi_i \, - \,m_i^2 \phi_i  \, + \, \frac{3 i\lambda}{2} \, \epsilon_{\mu\nu} \, \epsilon_{ijk}\, (\partial_\mu \phi_j) \, (\partial_\nu \phi_k) = 0
\label{eqm}
\eeq

\subsection{Finite Action Solutions to Equation of Motion}   
Obviously, no real, non-trivial solutions exist to these equations of motion. To obtain non-trivial solutions we must complexify the fields.  We could, in principle,  take one field complex, or  all three complex, either choice will render the equations of motion (\ref{eqm}) real in either case.  We find that taking one field complex does not lead to a non-singular solutions.  Hence we take the ansatz
\begin{eqnarray}
\phi_1 (r,\theta) & = & i \, f(r) \, \cos{\omega \theta} \nonumber \\
\phi_2 (r,\theta) & = & i \, f(r) \, \sin{\omega \theta} \nonumber \\
\phi_3 (r,\theta) & = & i \, g(r)
\label{ansatz}
\end{eqnarray}
where for periodicity, actually,  $\omega=N$ for some integer $N$.  
To  separate the $\theta$ dependence we must take $m_1=m_2=m$ and then we get the equations 

\begin{eqnarray}
 r^2 \, f'' \, + \, r \, f' \, - \, (m^2 r^2 + \omega^2) \, f \, + \, 3 \omega \lambda r f g' & = & 0\\
\label{phi1}
 r^2 \, g'' \, + \, r \, g' \, - \, m^2_3 r^2 \, g \, - \, 3 \omega \lambda r f f' & = & 0
\label{phi3}
\end{eqnarray}
where the prime on functions means differentiation with respect to $r$. Also we have suppressed the functional dependences of $f$ and $g$ on $r$.
We notice that the above equations simplify significantly if we take all particles to be massless i.e. $m_1 = m_2 = m_3 = 0$:
\begin{eqnarray}
 r^2 \, f'' \, + \, r \, f' \, - \,  \omega^2 \, f \, + \, 3 \omega \lambda r f g' & = & 0\\
\label{eqfr}
 r^2 \, g'' \, + \, r \, g' \, - \, 3 \omega \lambda r f f' & = & 0
\label{eqgr}
\end{eqnarray}
Equation (\ref{eqgr}) integrates directly as
\beq
 (r g')' =  \frac{3}{2} \omega \lambda  (f^2)'
\label{eqg1}
\eeq
which gives
\begin{eqnarray}
 g' & = & \frac{3}{2 r} \omega \lambda f^2 \, + \, \frac{c_1}{r}
\label{eqgp1'}
\end{eqnarray}
where $c_1$ is an arbitrary integration constant. Finite Euclidean action, after some algebra, requires $c_1 = 0$.  Thus we get  
\beq
 g' =  \frac{3}{2 r} \omega \lambda f^2 .
\label{eqg1'}
\eeq
Then using (\ref{eqg1'}) in (\ref{eqfr}) we have
\begin{eqnarray}
 r^2 \, f'' \, + \, r \, f' \, - \,  \omega^2 \, f \, + \, 3 \omega \lambda r f \left( \frac{3}{2 r} \omega \lambda \,f^2 \right)  & = & 0 \nonumber \\
\Rightarrow \qquad  r^2 \, f'' \, + \, r \, f' \, - \,  \omega^2 \, f \, + \, \frac{9}{2} \omega^2 \lambda^2 \, f^3 & = & 0
\label{eqf1}
\end{eqnarray}
Multiplying (\ref{eqf1}) by $f'$ and integrating and after some trivial algebra, gives
\beq
 (f')^2 = \frac{\omega^2\, f^2}{r^2} \left ( 1 \, - \, \frac{9}{2} \omega^2 \lambda^2\, f^2  \right).
\label{eqf11}
\eeq
This yields, after elementary integration,
\beq
 f \, = \, \frac{4}{3 \lambda} \, \frac{(r/r_0)^{\pm \omega}}{ \left((r/r_0)^{\pm 2 \omega}  + 1 \right)}
\eeq
where $r_0$ is effectively the integration constant. 
One can also check that for both $\pm \omega$ we get the same solution  for $f$:
\begin{eqnarray}
 f & = & \frac{4}{3 \lambda} \, \frac{1}{ \left((r/r_0)^{ \omega}  + (r/r_0)^{-\omega} \right)}\label{eqfp}
\label{eqf}
\end{eqnarray}
Using (\ref{eqfp}) in (\ref{eqg1'}) and integrating we get
%
%we have 
%
%\begin{eqnarray}
% g' & = & \frac{3}{2 r} \omega \lambda \left[ \frac{4\,(r/r_0)^{ \omega}}{3 \lambda\, \left( (r/r_0)^{ 2 \omega}  + 1 \right)} \right]^2 \nonumber \\
%\Rightarrow \qquad g' & = & \frac{8 c_3 \, \omega}{3 \lambda} \, \frac{(r/r_0)^{(2\omega - 1)}}{ \left((r/r_0)^{ 2 \omega}  + 1 \right)^2}
%\label{eqg'}
%\end{eqnarray}
%
%
%Integrating (\ref{eqg'}) we have 
\begin{eqnarray}
 g & = & \frac{-4}{3 \lambda} \,  \frac{1}{ \left((r/r_0)^{ 2 \omega}  + 1 \right)} \, + \, c
\label{eqg}
\end{eqnarray}
where $c$ is an integration constant.
%
%Also we have 
%
%\begin{eqnarray}
% f' & = & \frac{4 c_3 \, \omega}{3 \lambda} \, \frac{ \left((r/r_0)^{ 2 \omega}  - 1 \right) \, (r/r_0)^{(\omega - 1)}}{ \left((r/r_0)^{ 2 \omega}  + 1 \right)^2}
%\label{eqf'}
%\end{eqnarray}
Hence the field solutions to the equations of motion can be written as
\begin{eqnarray}
\phi_1 (r,\theta) & = & \frac{4i}{3 \lambda} \, \frac{(r/r_0)^{\omega}}{ \left((r/r_0)^{ 2 \omega}  + 1 \right)} \, \cos{\omega \theta} \nonumber \\
\phi_2 (r,\theta) & = &  \frac{4i}{3 \lambda} \, \frac{(r/r_0)^{\omega}}{ \left((r/r_0)^{ 2 \omega}  + 1 \right)} \, \sin{\omega \theta} \nonumber \\
\phi_3 (r,\theta) & = & - \, \frac{4i}{3 \lambda} \,  \frac{1}{ \left((r/r_0)^{ 2 \omega}  + 1 \right)} \, + \, ic
\label{phisol}
\end{eqnarray}

%%%%%%%%%%%%%%%%%%%%%%%%%%%%%%%%%%%%%%%%%%%%%%%%%%%%%%%%%%%%%%%%%%%%%%%%%%%%%%%%%%%%%%%%%%%%%%%%%%%%%%%%%%%%%%%%%%%%%%%%%%%%%%%%%%%%%%%%%%%%%%%%%%%%%%%%%%%%%%%%%%%%%%%%%%%%%%%%%%%%%%%

\subsection{The Euclidean action for our solution}
%The action (\ref{euclideanaction}) in polar coordinates becomes terms of our Ansatz (\ref{ansatz}) becomes 
%\begin{eqnarray}
% S_E & = & \frac{-1}{2} \, \int^\infty_0  r \, dr \, \int^{2\pi}_0 d\theta \, \left[ \left( \frac{\partial \phi_i}{ \partial r} \right)^2  
%\, + \, \frac{1}{r^2} \, \left( \frac{\partial \phi_i}{ \partial \theta } \right)^2 
%\, - \, 2 i \lambda \,\epsilon_{ijk} \, \frac{\phi_i}{r} \, \left( \frac{\partial \phi_j}{ \partial r} \right) \, \left( \frac{\partial \phi_k}{ \partial \theta} \right) \right]
%\label{polaraction}
%\end{eqnarray}
%
%In terms of our Ansatz (\ref{ansatz}), (\ref{polaraction}) becomes 
%
%\begin{eqnarray}
% S_E & = & \frac{-1}{2} \, \int^\infty_0  r \, dr \, \int^{2\pi}_0 d\theta \, \left[ - \left( f' \right)^2 \, - \, \left( g' \right)^2 \, - \,  \frac{\omega^2}{r^2} \,f^2 
%\, + \, \frac{2 \omega \, \lambda}{r} \, f^2 \, g' \, - \, \frac{2 \omega \, \lambda}{r} \, g \, f \, f' \right]
%\label{ansatzaction1}
%\end{eqnarray}
%
%Since (\ref{ansatzaction1}) is independent of $\theta$ so the integral over $\theta$ can be trivially done. Hence we get
In terms of our ansatz, \eqref{ansatz}, the Euclidean action \eqref{euclideanaction} becomes
\beq
 S_E  =  -\pi \, \int^\infty_0   dr \, \left[ r \, \left( f' \right)^2 \, + \, r \, \left( g' \right)^2 \, + \,  \frac{\omega^2}{r} \, f^2 
\, - \, 2 \omega \, \lambda \, f^2 \, g' \, + \, 2 \omega \, \lambda \, g \, f \, f' \right]
\label{ansatzaction}
\eeq
We first note that the action is independent of the constant that we could add to $g$ in Eqn. \eqref{eqg}, the terms involving $g'$ of course do not see the constant, and the final term changes by a total derivative, which integrates to zero given the boundary conditions  $f|_{r=0}=f|_{r=\infty}=0$.  Then using the equations of motion for $g'$ and for $f'$, Eqn. \eqref{eqg1'} and Eqn. \eqref{eqf1}, we find
\beq
S_E  =  -\pi \, \int^\infty_0   dr  \, r\left( \frac{\omega^2}{r^2} \, f^2 
\, - \, \frac{9 \omega^2 \, \lambda^2 }{2r^2}f^4\right).
\eeq
We could substitute the solution for $f$ directly into this expression and integrate, but there is a more elegant method.  We use Eqn. \eqref{eqf11} to insert unity into the integral
\bea
S_E  &=&  -\pi \, \int^\infty_0   dr  \,r\left( \frac{\omega^2}{r^2} \, f^2 \, \left( 1 - \, \frac{9  \, \lambda^2 }{2}f^2\right) \frac{f'}{\frac{\omega f}{r}\sqrt{1-\frac{9\lambda^2}{4}f^2}}\right)\nonumber\\
&=& \frac{-\pi\omega}{2} \, \int^\infty_0   dr  \,  \left( \frac{1 - \, \frac{9  \, \lambda^2 }{2}f^2}{\sqrt{1-\frac{9\lambda^2}{4}f^2}}\right)(f^2)'\nonumber\\
&=& 2\times  \frac{-\pi\omega}{2} \, \int^{\frac{4}{9\lambda^2}}_0   dx  \,  \left( \frac{1 - \, \frac{9  \, \lambda^2 }{2}x}{\sqrt{1-\frac{9\lambda^2}{4}x}}\right)
\eea
where we have used the fact that $x=f^2$ rises to its maximum value $f_{max}=\frac{4}{9\lambda^2}$ and then falls back down to zero, and thus we integrate only up to this value with the positive square root and multiply the result by 2. The integral is again elementary and yields
\beq
S_E=\frac{8\pi\omega}{27\lambda^2}.
\eeq

\section{Addition of a quartic potential }
\subsection{Minkowski solution}
We have observed in Eqn. \eqref{sumphi2} that
\beq
\sum_i\phi^2_i=\frac{4(m^2-\omega^2)}{9\lambda^2\omega^2}.
\eeq
Hence if we add the potential
\beq
V(\phi_i)= \gamma\left(\sum_i\phi^2_i-\frac{4(m^2-\omega^2)}{9\lambda^2\omega^2}\right)^2
\eeq
to the action, its contribution to the equations of motion
\beq
\frac{\partial V}{\partial\phi_i}=2\gamma\left(\sum_i\phi^2_i-\frac{4(m^2-\omega^2)}{9\lambda^2\omega^2}\right)2\phi_i
\eeq
 will exactly vanish for the solution that we have found.  Thus the full potential will correspond to the spontaneous symmetry breaking potential $V(\phi)$ in addition to the explicit symmetry breaking mass terms for the fields $(m^2/2)\phi_1^2$ and  $(m^2/2)\phi_2^2$.  The potential $V(\phi)$ will spontaneously break the original symmetry $SO(3)\to SO(2)$, giving rise to one massive scalar with $M^2=\frac{8\gamma(m^2-\omega^2)}{9\lambda^2\omega^2}$ and two massless scalar fields.  The explicit symmetry breaking terms preserve the $SO(2)$ symmetry, however, cause the putatively massless Goldstone bosons of the spontaneous symmetry breaking to become ``pseudo-Goldstone'' bosons of mass $m$.  For the notion that the ``pseudo-Goldstone'' boson fields are much lighter than the massive field, we should like to have $M>>m$, however, this is not at all required for our solutions to exist.
 
 \subsection{Euclidean solution}

We start with the observation that, the constant in Eqn. \eqref{phisol} for $\phi_3=ig$ is not at all determined, and does not affect the value of the euclidean action.  If we choose $c=2/3\lambda$ we find
\beq
g=\frac{-2}{3\lambda}\left(\frac{1-(r/r_0)^{2\omega}}{1+(r/r_0)^{2\omega}}\right),
\eeq
and then 
\beq
\sum_i\phi_i^2=-(f^2+g^2)=-\frac{16}{9\lambda^2}.
\eeq
Therefore, if we add the potential
\beq
V_E(\phi_i)=\gamma_E\left(\sum_i\phi_i^2+\frac{16}{9\lambda^2}\right)^2
\eeq
as in the Minkowski case, the contribution to the equations of motion will exactly vanish.  The potential added is not of the symmetry breaking type, all the fields become massive, with mass $M^2=4\gamma_E \frac{16}{9\lambda^2}$.  

\section{Conclusion}
We have studied a model of possible CP-violation where in the 1+1 dimensional analog, we find exact solitons of finite energy and exact instantons of finite Euclidean action.  The instantons could have an interpretation as exact solitons of a 2+1 dimensional theory, although the structure of our theory requires additional fields in higher dimensions.  Exact solitons in a somewhat related model,  were found a long time ago by Jackiw and Pi \cite{Jackiw:1990tz}.  The Jackiw-Pi model contains a Chern-Simons term which our interaction imitates, and a quartic interaction between the Schrodinger field, which we generate when we isolate the equation for say $f$, in Eqn. \eqref{eqf1}. The energy and the action of our solutions depends, as expected,  non-perturbatively on the coupling constant, hence we believe that these classical solutions will give rise to new non-perturbative contributions to CP-violation.    It is not clear what tunnelling our instanton solutions  describe.  The instanton solutions are established for the massless, potential free theory, however, they are also valid for the theory with a standard quartic self coupling between the fields, which are degenerate in mass.  There is no obvious meta-stable state whose decay is mediated by the instantons. 

The Minkowski solutions are of the $Q$-ball type, and for large $Q$, they owe their stability to the fact that the energy increases much slower than linearly for large charge.  Therefore they are energetically stable against disintegration into $Q$ perturbative, massive particles. Interestingly, even for small $Q$, our solitons have less energy than $Q$ perturbative, massive particles, $mQ$.  We then can imagine that the perturbative excitations are not stable, and should decay into $Q$-ball type solutions.  This kind of instability seems new, we are not aware of it in any other model.  With the addition of the quartic potential term of the symmetry breaking type, because of the ``pseudo-Goldstone'' mass terms, the potential has in fact exactly two discrete, degenerate vacua, $\phi_1=\phi_2=0$, $\phi_3=\pm \frac{2\sqrt{m^2-\omega^2}}{3\lambda\omega}$.  The fields of our $Q$-ball type soliton interpolate between the two vacua.  In principle, there should exist instantons which tunnel between the two vacua, however, we find no such instantons.  The instantons we find are for a modified theory which has a unique vacuum at $\phi_i=0$.

It would be interesting and important to generalize our results to a 3+1 dimensional model.

\maketitle

\section{ACKNOWLEDGEMENTS}

We thank Bhujyo Bhattacharya for useful discussions.  We thank NSERC, Canada for financial support. The visit of NC and RH to Montréal, where this work was completed, was possible due to fellowships from the Canadian Commonwealth Scholarship, we gratefully acknowledge the assistance.  This work was written up while MP was visiting Urjit Yajnik at IITBombay, Mumbai, India and the Inter-University Center for Astronomy and Astrophysics, Pune, India, their hospitality is also gratefully acknowledged.

%%%%%%%%%%%%%%%%%%%%  BIBLIO  %%%%%%%%%%%%%%%%%%%%%%%%%%%

\bibliographystyle{apsrev}
\bibliography{ref}

\begin{thebibliography}{26}
\expandafter\ifx\csname natexlab\endcsname\relax\def\natexlab#1{#1}\fi
\expandafter\ifx\csname bibnamefont\endcsname\relax
  \def\bibnamefont#1{#1}\fi
\expandafter\ifx\csname bibfnamefont\endcsname\relax
  \def\bibfnamefont#1{#1}\fi
\expandafter\ifx\csname citenamefont\endcsname\relax
  \def\citenamefont#1{#1}\fi
\expandafter\ifx\csname url\endcsname\relax
  \def\url#1{\texttt{#1}}\fi
\expandafter\ifx\csname urlprefix\endcsname\relax\def\urlprefix{URL }\fi
\providecommand{\bibinfo}[2]{#2}
\providecommand{\eprint}[2][]{\url{#2}}

\bibitem[{\citenamefont{Burgess and Moore}(2006)}]{burgess2006standard}
\bibinfo{author}{\bibfnamefont{C.}~\bibnamefont{Burgess}} \bibnamefont{and}
  \bibinfo{author}{\bibfnamefont{G.}~\bibnamefont{Moore}},
  \emph{\bibinfo{title}{The standard model: A primer}}
  (\bibinfo{publisher}{Cambridge University Press}, \bibinfo{year}{2006}).

\bibitem[{\citenamefont{Kobayashi and Maskawa}(1973)}]{kobayashi1973cp}
\bibinfo{author}{\bibfnamefont{M.}~\bibnamefont{Kobayashi}} \bibnamefont{and}
  \bibinfo{author}{\bibfnamefont{T.}~\bibnamefont{Maskawa}},
  \bibinfo{journal}{Progress of Theoretical Physics}
  \textbf{\bibinfo{volume}{49}}, \bibinfo{pages}{652} (\bibinfo{year}{1973}).

\bibitem[{\citenamefont{Sakharov}(1967)}]{sakharov1967pis}
\bibinfo{author}{\bibfnamefont{A.}~\bibnamefont{Sakharov}},
  \bibinfo{journal}{Sov. Phys. JETP Lett} \textbf{\bibinfo{volume}{5}},
  \bibinfo{pages}{24} (\bibinfo{year}{1967}).

\bibitem[{\citenamefont{Kuzmin}(1970)}]{Kuzmin:1970nx}
\bibinfo{author}{\bibfnamefont{V.~A.} \bibnamefont{Kuzmin}},
  \bibinfo{journal}{Pisma Zh. Eksp. Teor. Fiz.} \textbf{\bibinfo{volume}{12}},
  \bibinfo{pages}{335} (\bibinfo{year}{1970}).

\bibitem[{\citenamefont{Weinberg}(1979)}]{Weinberg:1979bt}
\bibinfo{author}{\bibfnamefont{S.}~\bibnamefont{Weinberg}},
  \bibinfo{journal}{Phys. Rev. Lett.} \textbf{\bibinfo{volume}{42}},
  \bibinfo{pages}{850} (\bibinfo{year}{1979}).

\bibitem[{\citenamefont{Coleman}(1979)}]{Coleman:1978ae}
\bibinfo{author}{\bibfnamefont{S.~R.} \bibnamefont{Coleman}},
  \bibinfo{journal}{Subnucl. Ser.} \textbf{\bibinfo{volume}{15}},
  \bibinfo{pages}{805} (\bibinfo{year}{1979}).

\bibitem[{\citenamefont{Padilla et~al.}(2011)\citenamefont{Padilla, Saffin, and
  Zhou}}]{Padilla:2010ir}
\bibinfo{author}{\bibfnamefont{A.}~\bibnamefont{Padilla}},
  \bibinfo{author}{\bibfnamefont{P.~M.} \bibnamefont{Saffin}},
  \bibnamefont{and} \bibinfo{author}{\bibfnamefont{S.-Y.} \bibnamefont{Zhou}},
  \bibinfo{journal}{Phys. Rev.} \textbf{\bibinfo{volume}{D83}},
  \bibinfo{pages}{045009} (\bibinfo{year}{2011}), \eprint{1008.0745}.

\bibitem[{\citenamefont{Padilla et~al.}(2010)\citenamefont{Padilla, Saffin, and
  Zhou}}]{Padilla:2010de}
\bibinfo{author}{\bibfnamefont{A.}~\bibnamefont{Padilla}},
  \bibinfo{author}{\bibfnamefont{P.~M.} \bibnamefont{Saffin}},
  \bibnamefont{and} \bibinfo{author}{\bibfnamefont{S.-Y.} \bibnamefont{Zhou}},
  \bibinfo{journal}{JHEP} \textbf{\bibinfo{volume}{12}}, \bibinfo{pages}{031}
  (\bibinfo{year}{2010}), \eprint{1007.5424}.

\bibitem[{\citenamefont{Streater and Wightman}(2000)}]{streater2000pct}
\bibinfo{author}{\bibfnamefont{R.~F.} \bibnamefont{Streater}} \bibnamefont{and}
  \bibinfo{author}{\bibfnamefont{A.~S.} \bibnamefont{Wightman}},
  \emph{\bibinfo{title}{PCT, spin and statistics, and all that}}
  (\bibinfo{publisher}{Princeton University Press}, \bibinfo{year}{2000}).

\bibitem[{\citenamefont{Durieux and Grossman}(2015)}]{Durieux:2015zwa}
\bibinfo{author}{\bibfnamefont{G.}~\bibnamefont{Durieux}} \bibnamefont{and}
  \bibinfo{author}{\bibfnamefont{Y.}~\bibnamefont{Grossman}},
  \bibinfo{journal}{Phys. Rev.} \textbf{\bibinfo{volume}{D92}},
  \bibinfo{pages}{076013} (\bibinfo{year}{2015}), \eprint{1508.03054}.

\bibitem[{\citenamefont{Aaij et~al.}(2014)}]{Aaij:2014qwa}
\bibinfo{author}{\bibfnamefont{R.}~\bibnamefont{Aaij}} \bibnamefont{et~al.}
  (\bibinfo{collaboration}{LHCb}), \bibinfo{journal}{JHEP}
  \textbf{\bibinfo{volume}{10}}, \bibinfo{pages}{005} (\bibinfo{year}{2014}),
  \eprint{1408.1299}.

\bibitem[{\citenamefont{Gronau and London}(1990)}]{gronau1990isospin}
\bibinfo{author}{\bibfnamefont{M.}~\bibnamefont{Gronau}} \bibnamefont{and}
  \bibinfo{author}{\bibfnamefont{D.}~\bibnamefont{London}},
  \bibinfo{journal}{Physical Review Letters} \textbf{\bibinfo{volume}{65}},
  \bibinfo{pages}{3381} (\bibinfo{year}{1990}).

\bibitem[{\citenamefont{Alves~Jr et~al.}(2008)\citenamefont{Alves~Jr,
  Andrade~Filho, Barbosa, Bediaga, Cernicchiaro, Guerrer, Lima~Jr, Machado,
  Magnin, Marujo et~al.}}]{alves2008lhcb}
\bibinfo{author}{\bibfnamefont{A.~A.} \bibnamefont{Alves~Jr}},
  \bibinfo{author}{\bibfnamefont{L.}~\bibnamefont{Andrade~Filho}},
  \bibinfo{author}{\bibfnamefont{A.}~\bibnamefont{Barbosa}},
  \bibinfo{author}{\bibfnamefont{I.}~\bibnamefont{Bediaga}},
  \bibinfo{author}{\bibfnamefont{G.}~\bibnamefont{Cernicchiaro}},
  \bibinfo{author}{\bibfnamefont{G.}~\bibnamefont{Guerrer}},
  \bibinfo{author}{\bibfnamefont{H.}~\bibnamefont{Lima~Jr}},
  \bibinfo{author}{\bibfnamefont{A.}~\bibnamefont{Machado}},
  \bibinfo{author}{\bibfnamefont{J.}~\bibnamefont{Magnin}},
  \bibinfo{author}{\bibfnamefont{F.}~\bibnamefont{Marujo}},
  \bibnamefont{et~al.}, \bibinfo{journal}{Journal of instrumentation}
  \textbf{\bibinfo{volume}{3}}, \bibinfo{pages}{S08005} (\bibinfo{year}{2008}).

\bibitem[{\citenamefont{Wess and Zumino}(1971)}]{wess1971consequences}
\bibinfo{author}{\bibfnamefont{J.}~\bibnamefont{Wess}} \bibnamefont{and}
  \bibinfo{author}{\bibfnamefont{B.}~\bibnamefont{Zumino}},
  \bibinfo{journal}{Physics Letters B} \textbf{\bibinfo{volume}{37}},
  \bibinfo{pages}{95} (\bibinfo{year}{1971}).

\bibitem[{\citenamefont{Witten}(1983{\natexlab{a}})}]{Witten:1983tw}
\bibinfo{author}{\bibfnamefont{E.}~\bibnamefont{Witten}},
  \bibinfo{journal}{Nucl. Phys.} \textbf{\bibinfo{volume}{B223}},
  \bibinfo{pages}{422} (\bibinfo{year}{1983}{\natexlab{a}}).

\bibitem[{\citenamefont{Novikov}(1982)}]{Novikov:1982ei}
\bibinfo{author}{\bibfnamefont{S.~P.} \bibnamefont{Novikov}},
  \bibinfo{journal}{Usp. Mat. Nauk} \textbf{\bibinfo{volume}{37N5}},
  \bibinfo{pages}{3} (\bibinfo{year}{1982}).

\bibitem[{\citenamefont{Skyrme}(1961)}]{skyrme1961non}
\bibinfo{author}{\bibfnamefont{T.~H.~R.} \bibnamefont{Skyrme}}, in
  \emph{\bibinfo{booktitle}{Proceedings of the Royal Society of London A:
  Mathematical, Physical and Engineering Sciences}} (\bibinfo{organization}{The
  Royal Society}, \bibinfo{year}{1961}), vol. \bibinfo{volume}{260}, pp.
  \bibinfo{pages}{127--138}.

\bibitem[{\citenamefont{Skyrme}(1962)}]{skyrme1962unified}
\bibinfo{author}{\bibfnamefont{T.~H.~R.} \bibnamefont{Skyrme}},
  \bibinfo{journal}{Nuclear Physics} \textbf{\bibinfo{volume}{31}},
  \bibinfo{pages}{556} (\bibinfo{year}{1962}).

\bibitem[{\citenamefont{Gisiger and Paranjape}(1998)}]{gisiger1998recent}
\bibinfo{author}{\bibfnamefont{T.}~\bibnamefont{Gisiger}} \bibnamefont{and}
  \bibinfo{author}{\bibfnamefont{M.~B.} \bibnamefont{Paranjape}},
  \bibinfo{journal}{Physics Reports} \textbf{\bibinfo{volume}{306}},
  \bibinfo{pages}{109} (\bibinfo{year}{1998}).

\bibitem[{\citenamefont{Witten}(1983{\natexlab{b}})}]{Witten:1983tx}
\bibinfo{author}{\bibfnamefont{E.}~\bibnamefont{Witten}},
  \bibinfo{journal}{Nucl. Phys.} \textbf{\bibinfo{volume}{B223}},
  \bibinfo{pages}{433} (\bibinfo{year}{1983}{\natexlab{b}}).

\bibitem[{\citenamefont{Witten}(1979)}]{witten1979baryons}
\bibinfo{author}{\bibfnamefont{E.}~\bibnamefont{Witten}},
  \bibinfo{journal}{Nuclear Physics B} \textbf{\bibinfo{volume}{160}},
  \bibinfo{pages}{57} (\bibinfo{year}{1979}).

\bibitem[{\citenamefont{Vilenkin and Shellard}(2000)}]{vilenkin2000cosmic}
\bibinfo{author}{\bibfnamefont{A.}~\bibnamefont{Vilenkin}} \bibnamefont{and}
  \bibinfo{author}{\bibfnamefont{E.~P.~S.} \bibnamefont{Shellard}},
  \emph{\bibinfo{title}{Cosmic strings and other topological defects}}
  (\bibinfo{publisher}{Cambridge University Press}, \bibinfo{year}{2000}).

\bibitem[{\citenamefont{Ablowitz and Clarkson}(1991)}]{ablowitz1991solitons}
\bibinfo{author}{\bibfnamefont{M.~J.} \bibnamefont{Ablowitz}} \bibnamefont{and}
  \bibinfo{author}{\bibfnamefont{P.~A.} \bibnamefont{Clarkson}},
  \emph{\bibinfo{title}{Solitons, nonlinear evolution equations and inverse
  scattering}}, vol. \bibinfo{volume}{149} (\bibinfo{publisher}{Cambridge
  university press}, \bibinfo{year}{1991}).

\bibitem[{\citenamefont{MacKenzie and Paranjape}(2001)}]{mackenzie2001q}
\bibinfo{author}{\bibfnamefont{R.~B.} \bibnamefont{MacKenzie}}
  \bibnamefont{and} \bibinfo{author}{\bibfnamefont{M.~B.}
  \bibnamefont{Paranjape}}, \bibinfo{journal}{Journal of High Energy Physics}
  \textbf{\bibinfo{volume}{2001}}, \bibinfo{pages}{003} (\bibinfo{year}{2001}).

\bibitem[{\citenamefont{Alexanian et~al.}(2008)\citenamefont{Alexanian,
  MacKenzie, Paranjape, and Ruel}}]{alexanian2008path}
\bibinfo{author}{\bibfnamefont{G.}~\bibnamefont{Alexanian}},
  \bibinfo{author}{\bibfnamefont{R.}~\bibnamefont{MacKenzie}},
  \bibinfo{author}{\bibfnamefont{M.}~\bibnamefont{Paranjape}},
  \bibnamefont{and} \bibinfo{author}{\bibfnamefont{J.}~\bibnamefont{Ruel}},
  \bibinfo{journal}{Physical Review D} \textbf{\bibinfo{volume}{77}},
  \bibinfo{pages}{105014} (\bibinfo{year}{2008}).

\bibitem[{\citenamefont{Jackiw and Pi}(1990)}]{Jackiw:1990tz}
\bibinfo{author}{\bibfnamefont{R.}~\bibnamefont{Jackiw}} \bibnamefont{and}
  \bibinfo{author}{\bibfnamefont{S.~Y.} \bibnamefont{Pi}},
  \bibinfo{journal}{Phys. Rev. Lett.} \textbf{\bibinfo{volume}{64}},
  \bibinfo{pages}{2969} (\bibinfo{year}{1990}).

\end{thebibliography}

\end{document}